\documentclass{article}

\usepackage{graphicx}
\usepackage{color}
\usepackage{amssymb}

\def\tf{\widetilde\vphi}
\def\tp{\widetilde\psi}
\def\cL{{\cal{L}}}
\def\pt{\partial_{t}}

\def\px{\partial_{x}}
\def\a{\alpha}
\def\b{\beta}

\def\de{\delta}

\def\ga{\gamma}

\def\k{\kappa}
\def\la{\lambda}
\def\om{\omega}
\def\s{\sigma}
\def\vphi{\varphi}
\def\vpsi{\psi}
\def\vth{\vartheta}

\def\px{\partial_{x}}
\def\pa{\partial}

\def\L{\mathcal{L}}

\def\fracor#1#2{({#1} / {#2} )}

\def\bea{\begin{eqnarray}}
\def\eea{\end{eqnarray}}
\def\beq{\begin{equation}}
\def\eeq{\end{equation}}
\def\lb{\label}

\def\eqref#1{(\ref{#1})}

\def\wt#1{\widetilde{#1}}

\def\^#1{\widehat{#1}}

\begin{document}

\title{ Speed selection for coupled wave equations}

\author{Mariano Cadoni\thanks{Dipartimento di Fisica, Universit\`a di
Cagliari \& INFN, Sezione di Cagliari; Cittadella Universitaria,
09042 Monserrato, Italy. E-mail: {\tt mariano.cadoni@ca.infn.it}}
\ and Giuseppe Gaeta\thanks{Dipartimento di Matematica,
Universit\`a degli Studi di Milano, via Saldini 50, 20133 Milano,
Italy. E-mail: {\tt giuseppe.gaeta@unimi.it}} {}\thanks{Partially
supported by MIUR-PRIN program under project 2010-JJ4KPA.}}

\bigskip


\maketitle

\begin{abstract} We discuss models for coupled wave equations
describing interacting fields, focusing on the speed of travelling wave solutions.
In particular,  we propose a general mechanism for selecting and tuning the
speed of the corresponding (multi-component)
travelling wave solutions under certain physical conditions.
A number of physical
models (molecular chains, coupled Josephson junctions,  propagation of kinks in chains of adsorbed atoms and domain walls)
are considered as examples.
\end{abstract}

\section{Introduction}

In a previous paper \cite{CDGspeed} dealing with a concrete
physical model -- more specifically, with the nonlinear dynamics
of the DNA macromolecule \cite{CDG1,CDG2} -- we have observed a
remarkable phenomenon. This is as follows: in the continuum limit
that model reduces to two coupled nonlinear wave equations for
different fields $\phi_{1,2}$; if the coupling is switched off,
each of the wave equations $E_{1,2}$ obeyed by the fields
$\phi_{1,2}$ is Lorentz invariant with \emph{different} limiting
speed, i.e. in particular admits travelling waves (solitons) with
any speed $c$ smaller than the limiting speed $c_{1,2}$. When the
full model, including the interaction, is considered, there is no
Lorentz invariance, and the travelling wave solutions admits only
a given speed  (see also \cite{CDG3}). Thus we have a selection
mechanism for the speed of travelling wave (TW) solutions (the latter
turn out to be also stable and thus physically relevant
\cite{CDG2,CDG3,DD,CDDG2,CDD}).

In the present paper we want to study if this ``speed selection
mechanism'' works  also  in a more general class of equations. We
will  answer to this in the positive, and actually the relevant
class of equations turns out to be rather ample. An important
point in this context is the fact that the speed selection
mechanism of the paper mentioned above  is
implemented using a constraint \cite{CDGspeed}. In general the constraint will
define a submanifold of the configuration space for the associated
dynamical system. The crucial requirement for our speed selection
mechanism to work is  that the constraint is natural, i.e the
associated  submanifold in invariant under the dynamics.  In
other words, if the initial data are chosen to lie on this
submanifold, the dynamics will take place entirely on it with no
need to introduce external forces to enforce the constraint.

We note that, apart from the theoretical interest, this question
also have potentially very relevant practical consequences.
In fact,  for instance, it would point out a way to have
transmission lines with physically determined speed for the
travelling wave packets.

As mentioned above, we will answer in the positive to the question
 of applicability of the speed selection mechanism to more
general equations. It turns out that, albeit we are mostly
interested in nonlinear equations, a speed selection mechanism is
also present in the case of linear systems. Note that for linear
equations the uncoupled equations have each a well definite speed,
so in this context the mechanism we study amounts to a change in
the value of the $n$ allowed speeds.

We will thus start, in Section \ref{sec:lin}, by considering
simple linear systems; in this case the speed selection follows
from  some trivial algebraic facts and  gives the possibility
of tuning the speed of TW just by changing the value
of an interaction parameter.

The general setting of this note will be as follows. We
investigate wave equations for $N$ fields $\phi^i (x,t),\,
i=1\ldots N$ in $1+1$  dimensions (one space dimension and time) described
by a Lagrangian, and we are specially interested in travelling
wave solutions.

The Lagrangian will be written as \beq \L \ = \ \sum_i \L_i \ + \
\L_{{\mathrm int}} \ , \eeq where the $\L_i $ is a Lagrangian for
each of the fields $ \phi^i$, and $\L_{{\mathrm int}}$ is the
interaction Lagrangian.

We will make ``minimal'' choices for $\L_i$ and $\L_{{\mathrm
int}}$, as we want to understand the phenomenon of speed selection
in the simplest possible terms. In particular, $\L_{{\mathrm
int}}$ will be made of a ``gradient interaction term'', coupling
the spatial gradient $\phi^i_x$ of different fields -- and
playing an essential role in our analysis -- and possibly of a
potential term $V (\phi_{i}) = V(\phi^1,...,\phi^N)$.

We will consider mainly  Lagrangians leading to hyperbolic wave
equations. In  appendix \ref{sec:parabolic}, we will extend our
considerations to parabolic, Schr\"odinger-like equations. Thus,
we take first a Lagrangian of the form \beq\label{eq:lag} \L \ = \
\frac{1}{2}\sum_{i=1}^{N}\left[\rho_i^2 \, \left(\pa_t \phi_{i}
\right)^{2} \, - \, \k_i^2 \, (\pa_x \phi_{i})^{2}\right] \ - \
\sum_{ij=1}^{N}\left[\ga_{ij} (\pa_x \phi_{i}) (\pa_x \phi_{j})
\right] \ - \ V(\phi_{i}) \ , \eeq where $\rho_i$, $\k_i$ are
constant, $\ga_{ij}$ are the components of  a constant $(N \times
N)$ matrix $\Gamma$, and the potential $V$ (which could be zero)
will be appropriately chosen below (see Sect.  \ref{sec:lin} and 
Sect. \ref{sec:nonlin}).
The fields and the constants are assumed to be real; note that the
speed of the  waves for the fields $\phi_i$ we get in the
decoupled case $\ga=V=0$ are \beq\lb{eq:speed} c_i \ = \ \k_i /
\rho_i \ . \eeq

 The matrix $\Gamma$ can be taken to be symmetric, and in
the present notation we can assume it has zero terms on the
diagonal, as the corresponding terms are represented by the $\k_i$
(so that $\Gamma$ only represents the interaction between
gradients of \emph{different} fields).

The Euler-Lagrange equations corresponding to the Lagrangian
\eqref{eq:lag} are\footnote{Here and below we move the field index
up and down for typographical convenience.} \beq\label{eq:ELgen}
\rho_i^2 \, \phi^i_{tt} \ - \ \k_i^2 \, \phi^i_{xx} \ - \ \ga_{ij}
\, \phi^j_{xx} \ = \ - \ (\pa V / \pa \phi^i ) \ . \eeq

It is appropriate to mention immediately some physical relevant
cases in which one meets Lagrangians of the type \eqref{eq:lag}.
This include e.g., beside the ``composite'' model of DNA dynamics
\cite{CDG1} mentioned above and the strictly related case of long
wavelength excitations in a chain of double pendulums (this also
applies to polyethylene \cite{CDDG}), the case of coupled
Josephson junctions \cite{KM,Par} and interaction between kinks in
coupled chains of adsorbed atoms \cite{KMad}. These cases will be
discussed later on as examples of our general mechanism; see
Sections \ref{sec:exalin} and \ref{sec:exanonlin}.

Our analysis would of course also apply to Lagrangians differing
from \eqref{eq:lag} only by boundary terms and total
differentials, e.g by a term
$\frac{1}{2}\sum_{ij=1}^{N}\sigma_{ij}\phi_{i}\px\phi_{j}$; we
focus on Lagrangians of the form \eqref{eq:lag} both for these
make the analysis rather transparent and for their physical
relevance.

Most of our discussion will be conducted in the simplest
nontrivial case, $N=2$; we will then discuss the generalization to
the arbitrary $N$ case, which will be in some case rather
immediate.

\section{The linear case}
\label{sec:lin}

As mentioned above, we start discussing the case of a quadratic
Lagrangian, hence of linear field (Euler-Lagrange) equations.

Moreover, we will at first consider the case $N=2$; we will then
write, for ease of notation, $\phi_{1}=\vphi$ and $\phi_{2}=\psi$;
there is only one nontrivial term in the matrix $\Gamma$, i.e.
$\ga_{12}$ (which, with a slight abuse of notation, we denote
simply by $\ga$). Thus we study the Lagrangian
\beq\label{eq:lag2d} \L \ = \ \frac{1}{2} \left[ \left( \rho_1^2
\vphi_t^2 \, + \, \rho_2^2 \psi_t^2 \right) \, - \, \left( \k_1^2
\vphi_x^2 \, + \, \k_2^2 \psi_x^2 \right) \right] \ - \ \ga \,
\vphi_x \, \psi_x \ - \ V(\vphi,\psi) \ . \eeq The Euler-Lagrange
field equations are then
\begin{eqnarray}
\rho_1^2 \, \vphi_{tt} \ - \ \k_1^2 \, \vphi_{xx} \ - \ \ga \,
\psi_{xx} &=&
- \, \fracor{\pa V }{ \pa \vphi}  \ ,\nonumber \\
\rho_2^2 \, \psi_{tt} \ - \ \k_2^2 \psi_{xx} \ - \ \ga \,
\vphi_{xx} &=& - \, \fracor{\pa V }{ \pa \psi } \ . \label{eq:EL}
\end{eqnarray}

In the simplest case, which we consider in this section, $V$ will
be a quadratic function of its arguments (including the case where
$V$ is trivial) and the equations \eqref{eq:EL} will hence be
linear.

We will write, for the sake of definiteness,
 \beq\lb{pot1}
V(\vphi,\psi) \ = \ \frac12 \left(\mu^2_1 \, \vphi^2 \ + \ 2 \la
\, \vphi \, \psi \ + \ \mu^2_2 \, \psi^2 \right) \ , \eeq where
$\mu_{1,2}^{2}$ are the masses of the non interacting fields
($\la=\ga=0$) and can hence be assumed to be positive.

Thus our Lagrangian is defined by three matrices $Q^{(t)}$,
$Q^{(x)}$, $Q^{(V)}$, given by \beq\label{eq:Qmat} Q^{(t)} \ = \
\pmatrix{\rho_1^2 & 0 \cr 0& \rho_2^2 \cr} \ ; \ \ Q^{(x)} \ = \
\pmatrix{ \k_1^2 & \ga \cr \ga & \k_2^2  \cr} \ ; \ \ Q^{(V)} \ =
\ \pmatrix{\mu_1^2 & \la \cr \la & \mu_2^2 \cr} \ ; \eeq we have
(returning for a moment to the notation with indices)
$$ \L \ = \ \frac12 \sum_{ij=1}^{N}¥\left[ \left( \phi^i_t \, Q^{(t)}_{ij} \,
\phi^j_t \right) \  - \ \left( \vphi^i_x \, Q^{(x)}_{ij} \,
\phi^j_x \right) \ - \ \left( \phi^i \, Q^{(V)}_{ij} \, \phi^j
\right) \right] \ . $$

The problem -- or better the source of interesting behavior --
lies in that, in general, the three matrices $Q^{(t)}$, $Q^{(x)}$
and $Q^{(V)}$ do not commute with each other. More specifically,
we have
\begin{eqnarray}
\left[ Q^{(t)} , Q^{(x)} \right] &=& (\rho_2^2 - \rho_1^2) \
\pmatrix{0 & - \ga \cr \ga & 0 \cr} \ ; \nonumber \\
\left[ Q^{(x)} , Q^{(V)} \right] &=& (\mu_1^2 - \mu_2^2) \
\pmatrix{0 & - \ga \cr \ga & 0 \cr} \ + \
(\k_2^2 - \k_1^2 ) \ \pmatrix{0 & - \la \cr \la & 0 \cr} \ ;
\label{eq:Qcomm} \\
\left[ Q^{(t)} , Q^{(V)} \right] &=& (\rho_2^2 - \rho_1^2 ) \
\pmatrix{0 & - \la \cr \la & 0 \cr} \nonumber\ . \end{eqnarray}

This lack of commutativity between the different matrices, and in
particular the fact we have $[ Q^{(t)} , Q^{(x)} ] \not= 0$, is
responsible for the breaking of the space-time Lorentz symmetry,
which is one of the sources of the unusual (and interesting)
features of the theory;  see also the discussion in Appendix
\ref{sec:group}.

\subsection{Travelling wave ansatz}
\lb{sect:twa}

If we look for travelling wave  solutions, i.e. for solutions
of the form \beq\label{eq:twa} \vphi (x,t) \ = \ \vphi (x \pm c t)
\ = \ \vphi (z)\ , \ \ \psi (x,t) \ = \ \vpsi (x \pm c t) \ = \
\vpsi (z) \ , \eeq we have then to study the equations
\begin{eqnarray}\lb{tweq}
\left( \rho_1^2  c^2 - \k_1^2 \right) \, \vphi_{zz} \ - \ \ga \,
\vpsi_{zz}
\ = \ - \, (\mu_1^2 \vphi + \la \vpsi ) \ , \nonumber \\
\left( \rho_2^2  c^2 - \k_2^2 \right) \, \vpsi_{zz} \ - \ \ga \,
\vphi_{zz} \ = \ - \, (\la \vphi + \mu_2^2 \vpsi ) \ .
\end{eqnarray}

These are immediately written in matrix form. Defining
$$ A \ = \ \pmatrix{[ \k_1^2 - \rho_1^2 c^2 ] &  \ga \cr
  \ga & [\k_2^2 - \rho_2^2 c^2] \cr} \ , \ \ B \ = \ \pmatrix{\mu_1^2
& \la \cr
\la  &  \mu_2^2 \cr} \ ; \ \Phi \ = \ \pmatrix{\vphi \cr \psi \cr} \ ,
$$
we recast the previous equation as \beq\label{eq:linTW} A \,
\Phi_{zz} \ = \ B \, \Phi \ . \eeq
If $\det (A) \not= 0$, i.e. if
$$ c^2 \ \not= \  \frac{1}{2 \rho_1^2 \rho_2^2 } \ \left[ (\k_1^2
\rho_2^2 + \k_2^2 \rho_1^2) \ \pm (\k_1^2 \rho_2^2 - \k_2^2 \rho_1^2)
\ \sqrt{1 + \frac{4 \ga^2 \rho_1^2 \rho_2^2}{(\k_1^2 \rho_2^2 -
\k_2^2 \rho_1^2)^2} } \ \right] \ , $$
we can invert $A$ and further recast \eqref{eq:linTW} as
\begin{eqnarray*} \Phi_{zz} &=& M \ \Phi \ , \\
M &:=& A^{-1} \ B \ = \ \mathcal{M} \ \pmatrix{ \ga \la - \k_2^2
\mu_1^2 + c^2 \mu_1^2 \rho_2^2 & - \k_2^2 \la + \ga \mu_2^2 + c^2 \la
\rho_2^2 \cr -\k_1^2 \la + \ga \mu_1^2 + c^2 \la \rho_1^2 & \ga \la -
\k_1^2 \mu_2^2 + c^2 \mu_2^2 \rho_1^2 \cr} \ ; \\
\mathcal{M} &=& \frac{1}{\ga^2 - (\k_1^2 - c^2 \rho_1^2) (\k_2^2 -
c^2 \rho_2^2)} \ , \\
\det (M) &=& \mathcal{M} \ (\la^2 \ - \ \mu_1^2 \, \mu_2^2) \ .
\end{eqnarray*}

One could then diagonalize the matrix $M$ by a change of basis
$\Phi = \Lambda \^\Phi$ so that in the new coordinates the equation
reads $\^\Phi_{zz} = \^M \^\Phi$ with $\^M = \Lambda^{-1} M
\Lambda = \mathrm{diag} (\de_1 (c) , \de_2 (c) )$ a diagonal
matrix, and now solve easily for $\^\Phi (z)$ and hence for $\Phi
(z) = \Lambda \^\Phi (z)$.  As the  functions $\de_i
(c)$ (which of course do also depend on the various parameters of
the equations) will in general satisfy $\de_1 (c) \not= \de_2
(c)$, we will have two normal modes  with  frequencies (in
$z$) $\omega^{(z)}_{1,2}$ satisfying
$(\omega_{1}^{(z)})^{2}=-\de_1 (c)$ and
$(\omega_{2}^{(z)})^{2}=-\de_2 (c)$. These relations correspond to
the two branches of the dispersion relations we will find in the
next section using the  Fourier transform.

Note that the situation is completely different in the $V=0$ case,
i.e. for $B=0$. In this case the equation \eqref{eq:linTW} reduces to
\beq\label{eq:linTWV0} A \, \Phi_{zz} \ = \ 0 \ , \eeq
which admits a solution if and only if $\det (A) = 0$, which should
be seen as a requirement on the speed $c$. More specifically, this
yields with simple algebra.
\beq\lb{kgh} c^2_{\pm} \ = \
\frac{1}{2} \left[ (c_{1}^2 +c_{2}^2) \ \pm \ (c_{1}^2 -c_{2}^2)
\ \sqrt{P} \right] \ , \eeq where we have written
$$ P \ =  1  \
+ \  4 \, \left[\frac{ \ga}{ \rho_1 \rho_2 (c_{1}^2 -c_{2}^2)}
\right]^{2} \ ; $$ here and below $c_{i} = | \k_i / \rho_i|$ is
the speed of decoupled waves defined in Eq. \eqref{eq:speed}.

The eigenvectors of the matrix $A$  corresponding to the
eigenvalues $c_{\pm}$ can be also easily computed; up to a
normalization factor, they are \beq\label{e4ab} \Phi_{\pm} \ = \
\left(\begin{array}{c}
\fracor{c_{1}}{c_{2}}\sqrt{(c_{2}^{2}-c_{\pm}^{2})/(c_{1}^{2}
-c_{\pm}^{2})}\\ 1 \end{array}\right) \ . \eeq \medskip

 It is worth stressing that equations (\ref{kgh}) and
(\ref{e4ab}) describe two remarkable features  of TWs in linear
coupled systems (\ref{tweq})

First, we have that the linear coupling between the $\vphi$ and
$\vpsi$ field in the wave equation (\ref{tweq}) forces a
synchronization of  the $\vphi$- and $\psi$-waves: they have to
propagate with the same speed, given either by $c_{+}$ or by
$c_{-}$. To see this let us first  diagonalize the kinetic   part
of the Lagrangian (\ref{eq:lag2d}). It is not a priori evident
that this is possible because the kinetic matrices  $Q^x$ e $Q^t$
in Eq. (\ref{eq:Qmat}) do not commute. However, one can show that
a $GL(2,\mathbb{R})$ transformation $S$ exists, which acting on
the vector $\Phi$ as $\Phi=S \widetilde \Phi$ diagonalizes the
Lagrangian (\ref{eq:lag2d}) with $V=0$: \beq\lb{e2b} \cL \ = \
\frac{1}{2}\ \left[\left(\frac{\pt \tf}{c_{+}} \right)^{2} \ + \
\left( \frac{\pt \tp}{c_{-}} \right)^{2} \ - \ (\px  \tf)^{2} \ -
\ (\px  \tp)^{2} \right] \ . \eeq We see that in the $\widetilde
\Phi$ frame  we have  two decoupled normal modes given by
(\ref{e4ab})   propagating respectively at speed
$c_{+}$ and $c_{-}$ given by Eq. (\ref{kgh}); however they
represent disjoint sectors, which cannot be superimposed.

Second, we can tune  the speeds of the $\vphi$ and $\vpsi$ waves
between zero and a maximum value by changing the coupling between
the two fields, which is parametrized by the (positive) ratio $r$,
\beq\lb{ub} 0 \ \le \ r \ := \ \left( \frac{\ga}{\rho_{1}
\rho_{2}} \right)^2 \ \le \ c_{1}^{2} \, c_{2}^{2}  \ := \ r_{max}
\ . \eeq In particular, let us start from the decoupled case; here
$r=0$ and we get  $c_{+}=c_{1}$ and $c_{-}=c_{2}$. If we now
increase $r$, then the value of $c_{+}$  increases  whereas that
of $c_{-}$ decreases. When  $r\to r_{max}$, $c_{+}$ tends to its
maximum value whereas  $c_{-}\to 0$. Notice that for $r>r_{max}$,
$c_{-}^{2}$ become negative, i.e  we have an imaginary speed for
the harmonic mode; this detects an instability of the system.
Notice also that Eq. (\ref{ub}) defines a general bound on the
coupling parameters, which can be also generalized to the case of
a non vanishing potential (\ref{pot1}).

\subsection{Fourier transform}
\lb{sect:ft}

To solve the general linear case we well make use of the linear
character of the field equations (\ref{eq:EL}), and pass to
consider the Fourier transforms $\^f(q,\om)$ and $\^g(q,\om)$ for,
respectively, the fields $\vphi (x,t)$ and $\psi (x,t)$; the
equations \eqref{eq:EL} are then recast as
\begin{eqnarray}
& & - \left( \rho_1^2 \om^2 - \k_1^2 q^2 \right) \, \^f \  + \ \ga
\, q^2 \, \^g \ + \
(\mu_1^2 \^f + \la \^g ) \ = \ 0 \ , \nonumber \\
& & - \left( \rho_2^2 \om^2  - \k_2^2 q^2 \right) \, \^g \  + \
\ga \, q^2 \, \^f \ + \ (\la \^f + \mu^2_2 \^g ) \ = \ 0 \ .
\label{eq:ELfou}
\end{eqnarray} These are promptly rewritten: defining now
$$ M \ = \ \pmatrix{[\rho_1^2 \om^2 - \k_1^2 q^2 - \mu^2_1] &  - \ga q^2
- \la \cr  - \ga q^2 - \la & [\rho_2^2 \om^2 - \k_2^2 q^2 -
\mu^2_2] \cr} \ , \ \ \^F \ = \ \pmatrix{\^f \cr \^g \cr} \ , $$
equations \eqref{eq:ELfou} read \beq M \, \^F \ = \ 0 \ . \eeq
Again we must require the vanishing of a determinant, i.e. $\det
(M) = 0$; this will now give a relation between $\om^2$ and $q^2$,
i.e. we will get some \emph{dispersion relations}  (DRs).

More precisely, these read  \beq\label{eq:DRgen} \om_\pm^2 \ = \
\frac{1}{2} \left[ (c_{1}^2 +c_{2}^2)q^{2}+ (u_{1}^2 +u_{2}^2) \pm
\ \sqrt{P} \ \right] \ , \eeq where we have written
$$ P \ = \left[ (c_{1}^2 -c_{2}^2)q^{2}+ (u_{1}^2
-u_{2}^2)\right]^{2} \ +\  4 \, \frac{ (\ga q^{2}  + \la)^{2}}{
\rho_1^2 \rho_2^2} \ , $$
 with $u_{i}=\mu_{i}/\rho_{i}$ and $c_{i}$ as above.

For a general nontrivial potential, we have two different DRs,
which will also determine the phase  velocity  $v^p_\pm = d\om_\pm/dq$ of
the  wave.
In this generic case the DR have a rather involved form. They take
a simpler form for some particular or limiting cases.

For a vanishing potential i.e $\mu_{1}=\mu_{2}=\la=0$, one can
easily check that the DR (\ref{eq:DRgen}) become linear (acoustic)
and that  $d\om_\pm /dq=c_{\pm}$ with $c_{\pm}$ given by
(\ref{kgh}). This case describes also the high energy limit  $q\to
\infty$ of the DR (\ref{eq:DRgen}), where the potential can be
neglected.

In the low-energy limit $q\to 0$ the DR (\ref{eq:DRgen}) describe
two optical branches, whose
analytic expression can be derived expanding
(\ref{eq:DRgen})  near $q=0$.

As expected, the DR become simple also in the decoupling limit
$\ga=\la=0$. In this case, we have two branches with optical DR
$\om^{2}_{12}= c^{2}_{12} q^{2}+ u_{12}^{2}$. Another interesting
particular case  is $c_{1}=c_{2},\, u_{1}=u_{2}$. We obtain  also
in this case two optical branches: \beq\lb{top}
\om_\pm^{2}=\left(c_{1}^{2}\pm
\frac{\ga}{\rho_{1}\rho_{2}}\right)q^{2}+u_{1}^{2}  \pm
\frac{\la}{\rho_{1}\rho_{2}} \eeq

The synchronization and  tuning effect discussed  in Sect.
(\ref{sect:twa}) for the TW speed $c$  applies also to the phase
speed $d\om_\pm /dq$. Now synchronization simply means that a
given phase of the $\vphi$ and $\psi$ wave must propagate at the
same speed, whereas tuning means that we can change the phase
velocity by acting on the coupling parameters of the two fields.
In the case of acoustic and optical dispersion relations  this
tuning is very simple. In the acoustic case,  the group and the
phase speed are the same, and thus it has been already described
in Sect. (\ref{sect:twa}). In the optical case we have instead
$d\om_\pm /dq= c_{\pm} q/\om_\pm $, so that the phase velocity
becomes zero whenever $c_{\pm}=0$ and for a given phase, grows
monotonically with $c_{\pm}$.

\subsection{The N-fields case and Lorentz symmetry}
\lb{sect:nfg}

The discussion conducted above is readily generalized to the case
of $N$ fields. This is specially transparent in terms of the 
travelling wave ansatz (TWA):
in this case we get again a matrix equation, but the matrix   is
now an $(N \times N)$ matrix. Solving the eigenvalue problem we
get in general $N$ normal modes.

In terms of the Fourier transform, we will get $N$ branches of the
dispersion relations.

Similarly to the $N=2$ case, for a vanishing potential we will
have  $N$ determinations of the allowed speeds $c$ and we can
change the values of the these speeds  (for given $\k_{i},\,
\rho_{i}$ and hence ``uncoupled speeds'' $c_i = \k_i / \rho_i$) by
acting on the coupling parameters  $\ga_{ij}$ of the model.

Let us now briefly comment on the Lorentz symmetry of our
two-dimensional field theory (\ref{eq:lag}). Space-time Lorentz
symmetry is explicitly broken in the Lagrangian (\ref{eq:lag}).
This breaking has two sources. The first is the presence of
several different limit speeds $c_{i}$; the breaking of the
Lorentz symmetry in this case is expressed by the non
commutativity of the matrices $Q^{(x)}$ and $Q^{(t)}$ in Eq.
(\ref{eq:Qmat}). The second source is the non covariance of the
kinetic coupling term, which contains  the space but not the time
derivative. On the other hand, by implementing the TWA
or by solving  the field equations in Fourier space the relevant
matrices   are not the  individual $Q^{(x)}$ and $Q^{(t)}$ but a
linear combination of them.

In Fourier space the breaking of the  Lorentz symmetry is  evident
from the form of the dispersion relations (\ref{eq:DRgen}). They
are invariant under Lorentz boosts only when $c_{1}=c_{2}$ and
$\ga=0$. Despite of the explicit breaking of the Lorentz symmetry,
it is quite evident that some remnant of it survives in the  field
theory (\ref{eq:lag}). This is in particular evident in the
massless case where the dispersion relation remain linear, the
Lagrangian can be diagonalized  describing  decoupled sectors  and
the presence of the coupling term $\ga$ just changes the values of
the two speeds $c_{12}$. We will discuss this remnant Lorentz
symmetry and related group theoretical aspects in Appendix
\ref{sec:group}.

\section{The nonlinear case}
\label{sec:nonlin}

Our discussion in Section \ref{sec:lin} makes a substantial use of
the linearity of field equations; thus it cannot be extended to
the nonlinear case.

In the nonlinear case, the TWA produces a system of nonlinear
ODEs, and TW solutions are obtained as
solutions $\phi_{i}(x \pm c t)$ of these nonlinear ODEs, which can be
considered as describing an equivalent mechanical system;  this
procedure  leaves  (in general, see below) the TW speed
completely undetermined.

This is particularly evident when when we start from a Lagrangian
which is Lorentz invariant: then the TW solution must be a
function of the relativistic gamma factor, $\phi_{i}=\phi_{i}((x\pm
ct)/\sqrt{1-c/\bar c})$ (where $\bar c$ is the limit speed). The
wave speed $c$ can be therefore arbitrarily changed in the range
$[0,\bar c]$ by a Lorentz boost. Thus any speed selection
mechanism  in the nonlinear case must necessarily start from a
Lagrangian in which the Lorenz invariance is explicitly broken to
remove this degeneracy in $c$.

Any speed selection mechanism must  constrain the dynamics of the
equivalent mechanical system to happen in a submanifold  of the
configuration space in which $c$ is fixed. Moreover, if the
submanifold corresponds to a \emph{natural constraint}, i.e. is
invariant under the dynamics, the constraint can also be realized
as a selection on the initial data. A simple  realization of such
a mechanism has been proposed in Ref. \cite{CDGspeed}. Here we
discuss in detail its dynamical implications, in particular
existence of invariant submanifolds and the naturalness of the
constraints that can be used to  select the speed of TW solutions.
Our main goal is obviously the generalization of the results of
Ref. \cite{CDGspeed}.

We are now considering a Lagrangian of the form \eqref{eq:lag}
with a generic (analytic) potential $V(\phi_{i})$, and Euler-Lagrange
equations \eqref{eq:ELgen}. In the case $N=2$ these reduce to
\eqref{eq:lag2d} and \eqref{eq:EL} respectively.

Note that the kinetic part of the Lagrangian -- and hence the
second order terms in the Euler-Lagrange equations -- is still characterized
by the two matrices $Q^{(t)}$ and $Q^{(x)}$, which in general do
not commute. This non commutativity expresses the breaking of the
space-time Lorentz symmetry at the Lagrangian level. On the other
hand, as already noticed in the discussion of the linear case,
after using the TWA (or passing to Fourier space) the relevant
matrix will be  a linear combination (with coefficients
depending on the speed of the TW) of $Q^{(t)}$ and $Q^{(x)}$.

With the travelling wave ansatz \eqref{eq:twa}, we are led to
consider the equations \beq (\rho_i^2  c^2 -  \k_i^2 ) \,
\phi^i_{zz} \ - \ \ga_{ij} \, \phi^j_{zz} \ = \  - \ \fracor{\pa V
}{ \pa \phi^i} \ . \eeq

\subsection{The two-dimensional case}
\lb{sect:tdc}

We will again start by considering the case $N=2$; with the
notation introduced above for the linear case, we are thus dealing
with the equations
\begin{eqnarray}
(\k_1^2  - \rho_1^2  c^2) \, \vphi_{zz} \ + \ \ga \, \psi_{zz} &=&
\fracor{\pa V}{ \pa \vphi} \ := \ f (\vphi,\psi )\
 \ , \nonumber \\
(\k_2^2  - \rho_2^2 c^2) \, \psi_{zz} \ + \ \ga \, \vphi_{zz} &=&
\fracor{\pa V}{ \pa \psi} \ := \ g (\vphi,\psi) \ .
\label{eq:ELNL2D}
\end{eqnarray}

These are rewritten in matrix form as
$$ M \, \Phi_{zz} \ = \ F \ , $$
having of course written
$$ M \ = \ \pmatrix{(\k_1^2  - \rho_1^2  c^2) & \ga \cr \ga & (\k_2^2
 - \rho_2^2  c^2)
\cr} \ := \ \pmatrix{m_1 & \ga \cr \ga & m_2 \cr} \ ; \ \ \ F =
\pmatrix{f \cr g \cr} \ . $$
 This matrix $M$ combines the $Q^{(x)}$ and $Q^{(t)}$.

Provided $\det (M) \not= 0$, i.e.  $\ga^2 \not= m_1 m_2$, we can
now
consider a linear change of coordinates in the fields space,
$$ \phi^i \ = \ A^i_{\ j} \ \eta^j \ , $$ which diagonalizes
\eqref{eq:ELNL2D}.\footnote{With $\wt{V} (\eta) = V [\phi
(\eta)]$, the latter reads $ M A \eta_{zz} = (A^{-1})^T  (\pa
\wt{V} / \pa \eta )$, so that we have to look for $A$ such that
$A^T M A$ is diagonal.}

We can actually ask more, i.e. that $A^T M A = I$. This is
obtained e.g. for
$$ A \ = \ \pmatrix{\a & - \sqrt{q (\a)}/\sqrt{m_2 (m_1 m_2 - \ga^2)}
\cr
- [\a \ga + \sqrt{q (\a )} ]/m_2 &
[\a \ga^2 - m_1 m_2 \a + \ga \sqrt{q(\a)}]/[m_2 \sqrt{m_1 m_2 -
\ga^2} \cr} \ , $$ where $\a$ is a free parameter and
$$ q(\a) \ := \ m_2 + (\ga^2 - m_1 m_2) \a^2 \ . $$

In terms of the fields $\eta$, our equations \eqref{eq:ELNL2D}
read now simply
$$ \frac{d^2 \eta_i}{d z^2} \ = \ - \frac{\pa W}{\pa \eta_i}
\ ; $$ thus the dynamics of the system is described by the motion
of a particle of unit mass in the effective potential
$W (\eta) := V [\phi (\eta)]$.

In dealing with PDEs one should also specify a function space for
the search of solutions. In view of the physical meaning of the
equations, it is natural to look for solutions of finite energy.
In the case of a natural Lagrangian like \eqref{eq:lag}, this
means that the solutions $\vphi (x,t)$ should go to a minimum of
the potential for $x \to \pm \infty$; once we proceed to reduction
to a system of ODEs via the TWA, this means that we look for
solutions which go to a minimum of $V$ (we stress the condition is
on $V$, \emph{not} on $W$) for $z \to \pm \infty$.

It is easily seen that if minima of $V$ correspond to minima of
the effective potential $W$, these solutions are necessarily
trivial. If, on the other hand, minima of $V$ correspond to maxima
of $W$, nontrivial solutions with the prescribed asymptotic
behavior can exist; moreover they can either be doubly asymptotic
to the same local minimum of $V$, or connect two distinct local
minima, corresponding to the same energy level.

Thus the existence of TW solutions with the relevant behavior
depends on the sign of $V/W$, i.e. on the sign of the
determinant of the Jacobian matrix $A$. This in turn will depend
on the value of $c^2$.  (The discussion of concrete examples will
 better clarify this point.)

Provided this condition is satisfied, we are thus searching for
solutions to our equivalent mechanical system with prescribed
limit conditions for $\eta (z)$ and $\eta_z (z)$ at $z \to \pm
\infty$. In particular, we require $$ \lim_{z \to \pm \infty}
\eta_z (z) \ = \ 0 \ ; \ \ \ \lim_{z \to \pm \infty} \eta_z (z) \
= \ \eta_\pm $$ with $\eta_\pm$ corresponding to (necessarily
degenerate) local minima of the potential $V$.

Such solutions are in general not unique (nor stable, even in the
realm of solutions satisfying the same limit conditions), as
immediately shown by the example of a doubly periodic potential,
i.e. of a dynamical system on the torus \cite{Rab,BR1,BR2}.

The situation -- and hence our analysis -- is greatly simplified
if there are some one-dimensional invariant submanifolds
connecting two degenerate minima. Needless to say, this is a
non-generic feature, and in general the existence of such
invariant submanifolds is related to the presence of some symmetry
in the potential\footnote{Strictly speaking, we are here concerned
with symmetries of the \emph{effective} potential; these will
however correspond to symmetries of the original physical
potential, as the transformation mapping the latter into the
former is smooth (and actually linear).}. This could be a Noether
symmetry, guaranteeing the presence of a conserved quantity and
hence the reduction to a lower dimensional effective dynamics; or
even a discrete symmetries -- such as a reflection symmetry -- in
the potential. More generally it suffices to have a conserved or
even a conditionally conserved quantity \cite{SL}; the existence
of this is implied  e.g. by the presence of a reflection symmetry.

Albeit in general a conditionally conserved quantity and hence a
low-di\-men\-sio\-nal invariant manifold is not necessarily
associated to a symmetry, in practice this is most often the case.

In this paper we will limit our considerations to the case in
which the system has a reflection symmetry, so that  we are
guaranteed of the existence of a low-dimensional invariant
submanifolds.
We have now to distinguish between three possible cases:

\begin{itemize}
\item[{\bf 1)}] The system is fully invariant under Lorentz
symmetry, i.e. we have $c_{1}=c_{2}=\bar c$ and $\ga=0$ in Eq.
\eqref{eq:ELNL2D}. In this case the invariant submanifold exists
for every value of the soliton  speed $c$ below the limit speed
$\bar c$. We do not have selection of the soliton speed. The
soliton speed is fully protected by the Lorentz symmetry. We will
give   examples of this situation in
Sects. \ref{sec:adatoms} and \ref{sec:Katsura}.

\item[{\bf 2)}] Lorentz symmetry is broken by $\ga\neq0$ but we
still have a single limit speed, i.e $c_{1}=c_{2}=\bar c$. Also in
this case the invariant submanifold exists for every value of the
soliton speed $c$ below a limit speed and the soliton speed is
protected by the Lorentz symmetry but the limit speed is changed
by the kinetic interaction term.  This is fully consistent with
the discussion  of the linear case of Sect. \ref{sec:lin}, where
we have shown that the kinetic  interaction term changes the wave
speed according to  Eq. \eqref{kgh}. This situation is realized in
an example discussed in Sect. \ref{sec:adatoms}.

\item[{\bf 3)}] Lorentz symmetry is broken by   $c_{1}\neq
c_{2}$. In this case the soliton speed is not protected by
the Lorentz symmetry. This is the case in which we can have speed
selection for the soliton; the invariant submanifold exists only
for selected values  of $c$. This situation is met in the examples
discussed in Sections \ref{sec:DPC}, \ref{sec:CJJ} and
\ref{sec:Katsura}.
\end{itemize}

Summarizing, a sufficient condition for the speed
selection mechanism  considered here to work  is
\emph{the presence of a reflection symmetry defining invariant
submanifolds connecting two degenerate minima of the potential $W$
existing only for specific values of the soliton speed.} Whereas a
necessary condition  for  this speed selection mechanism to
work is \emph{an explicit breaking of the Lorentz symmetry at the
Lagrangian level trough the presence of  different
limiting speeds.}
\bigskip

The practical implementation of the  speed selection mechanism
requires the  determination  of the invariant manifold  associated
with  the reflection symmetry. This is a rather involved problem,
which we will  tackle using  an ansatz. Obviously the manifold we
are looking for must be invariant not only under the dynamics but
also under the action of the  reflection  transformation. Because
the kinetic part of the field equations (\ref{eq:ELNL2D}) is
linear the most natural ansatz for determining the  invariant
manifold is a linear equation involving the fields $\vphi$ and
$\psi$. This will be our  choice for the examples described in
Sects.  \ref{sec:DPC}, \ref{sec:adatoms}, \ref{sec:CJJ}.
It is also possible to
search for invariant manifolds using  non linear invariant functions
involving the fields $\vphi$ and $\psi$. This will be our choice
for the  system described in Sect.
\ref{sec:Katsura}.

As a  final step in the practical implementation of the speed
selection mechanism we need   to restrict the dynamics on  the
invariant submanifold.  This can be achieved in two different
ways. A first  way is through a constraint simply given by the
ansatz we have used to determine the invariant manifold. Another
way,  making use of the fact we have a dynamically invariant
manifold and hence a \emph{natural constraint}, is to
suitably select the initial conditions,  i.e. choose them
(position and velocity) along the chosen invariant manifold.

\subsection{The $N$-dimensional case}

The analysis can be conducted along the same lines also in the
general $N$-dimensional case. Needless to say, even the
preliminary step of diagonalizing the matrix $M$ may be a
substantial problem for high dimension; but we may in principles
proceed along the same lines.

Note that for dimension $N$ a Noether symmetry will reduce the
dynamics to a problem in dimension $N-1$; thus we need $N-1$
Noether symmetries in involution in order to reduce our problem to
a one-dimensional one. Similarly, the presence of a reflection
symmetry will in general only guarantee the existence of an
invariant submanifold of dimension $N-1$ (i.e. of codimension 1);
thus we will need $N-1$ reflection symmetries across planes with a
nontrivial intersection to be sure of the existence of an
invariant one-dimensional manifold. Note this is e.g. the case if
the potential $V(\phi_{i})$ actually depends only on the squares
$\phi_i^2$ of the fields, or at least  of $N-1$ among them.

Another situation which guarantees the existence of invariant
one-dimensional submanifolds is that of a separable potential,
$V(\phi_1,...,\phi_N) = V_1(\phi_1) + ... + V_N(\phi_N) $,
provided each $V_k (\phi_{k}))$ has a minimum in $\phi_{k}=0$; in this case we do
not have to require reflection symmetry of the $V_k$.

One can of course also have a combination of the two situations
mentioned above, i.e. a potential which is separable in potentials
depending on several groups of field variables, each of them
having suitable (continuous or discrete) symmetries.

Provided the potential has a sufficient degree of symmetry or
separability, we are reduced to a one-dimensional analysis and we
can proceed  substantially as in the $N=2$ case above.

It is also possible that no symmetry is present, but that for
specific values of the speed some one-dimensional invariant
submanifold is present
We expect however that this becomes increasingly unlikely -- and
anyway that it would be increasingly difficult to determine the
allowed $c$ in concrete terms -- with increasing dimension $N$.

\section{Examples. Linear equations}
\label{sec:exalin}

In this section we  briefly apply our speed selection
mechanism developed for the linear case to two examples. These
describe the region of linear dynamics  for some of the non
linear systems to be considered in full in the next
Section.

\subsection{Kinks in coupled chains of adsorbed atoms}

In the situation to be considered in Section \ref{sec:adatoms},
the linearized equations at $(\vphi,\vpsi) = (0,0)$ read
(setting
$\beta=\ga$ to comply with our nomenclature of coupling constants for
the linear case)
\begin{eqnarray}
\vphi_{tt} \ - \ \vphi_{xx} &=& - \vphi \ - \ \a \, (\vphi -
\vpsi) \ + \ \ga \, \vpsi_{xx} \nonumber \\
\vpsi_{tt} \ - \ \vpsi_{xx} &=& - \vpsi \ - \ \a \, (\vpsi -
\vphi) \ + \ \ga \, \vphi_{xx} \ . \end{eqnarray}

This linear system has exactly the form considered in Sect.
\ref{sec:lin}, with special values of the parameters:
\beq\lb{fde}
\rho_{1}=\rho_{2}=\k_{1}=\k_{2}=1,\quad
\mu_{1}^{2}= \mu_{2}^{2}=1+\alpha,\quad \la= -\alpha
\eeq

Here $c_{1}=c_{2} = \bar{c} =1$, $u_{1}=u_{2}$;  thus Eq.
(\ref{top}) applies, giving the  optical DRs
$$ \om_\pm^{2}\ =\ (1\pm\ga) \, q^{2} \ + \ (1+\a)  \mp \a \ , $$ from which we get the
phase speeds for the two branches
$$  c_- (q) \ = \ \frac{1 - \ga}{\sqrt{1 + 2 \a + (1 - \ga) q^2}} \ q
\ ; \ \ c_+ (q) \ = \ \frac{1+\ga}{\sqrt{1 + (1 + \ga) q^2}} \ q \ .
$$
Note that as a consequence of Eq. (\ref{ub}) we must have
$|\ga|\le1$ and, as expected, for $\ga=1$ ($\ga=-1$) only the
$\om_+$  ( $\om_-$) branch is present, whereas the $\om_-$
($\om_+$)  branch disappears.

In the low-energy limit we get
$$ c_- (q) \simeq \frac{1-\ga}{\sqrt{1 + 2 \a}} \,  q \ , \ \ c_+
(q) \simeq (1 + \ga) \,  q \ ; $$ and in the high-energy limit
$$ c_- (q) \simeq \sqrt{1 - \ga} \ , \ \ c_+ (q) \simeq \sqrt{1 +
\ga} \ . $$

\subsection{Coupled Josephson junctions}

In the situation to be considered in Section \ref{sec:CJJ}, the
linearized equations at $(\vphi,\vpsi) = (0,0)$ read  (setting
$\a=-\ga$, again to comply with our general notation)
\begin{eqnarray}
\vphi_{tt} \ - \ \vphi_{xx} &=& - \vphi \ + \ \ga \, \vpsi_{xx}
\nonumber \\
\mu^2 \, \vpsi_{tt} \ - \ \vpsi_{xx} &=& - \, \nu^2 \, \vpsi \ + \
\ga \, \vphi_{xx} \ . \end{eqnarray}

This linear system has  the form considered in Sect.
\ref{sec:lin} with special values of the parameters:
\beq\lb{fde1}
\rho_{1}=\k_{1}=\k_{2}=1,\quad \rho_{2}=\mu,\quad
\mu_{1}=1,\quad \mu_{2}=\nu,\quad \la= 0
\eeq

The dispersion relations (\ref{eq:DRgen}) give
$$ \om^2_\pm \ = \ \frac{(\mu^2 + \nu^2 + (1
+ \mu^2) q^2 )}{2 \mu^2} \ \left[ 1 \ \pm \ \sqrt{1 - 4 \mu^2
\frac{\nu^2 + (1 + \nu^2 ) q^2 + (1 - \ga^2) q^4}{\left(\mu^2 + \nu^2 +
(1 + \mu^2) q^2\right)^{2}} } \right] \ . $$

The low-energy limit ($q \simeq 0$) for the speeds $ c_\pm = d
\om_\pm / d q$ are given by
$$ c_- (q) \ \simeq \ \frac{1}{\mu \nu} \ q \ + \ O(q^3 ) \ ; \ \
c_+ (q) \ \simeq \ q \ + \ O(q^3 ) \ . $$ In the high-energy limit
($q \to \infty$) both speeds go to finite limits; the explicit
expressions can be computed but are rather cumbersome and thus not
reported.

\section{Examples. Nonlinear equations}
\label{sec:exanonlin}

In this section we apply our TW  speed selection mechanism
developed  for the non linear case to  several examples.
For all of these examples we look for an invariant submanifold  of
the dynamics for the associated  mechanical system.

\subsection{Double pendulums chains}
\label{sec:DPC}

The speed selection mechanism for  TW solutions was originally
proposed \cite{CDGspeed} in the context of mesoscopic models for non
linear DNA  torsion dynamics
\cite{CDG1,CDG2,CDG3}. In particular, this was in studying
the ``composite Y model'' \cite{CDG1} of DNA, in which the state of
DNA is described by two angular variables $\vphi (x,t)$ and $\vth
(x,t)$, which can be thought as describing (in the long wavelength
limit, and thus using a continuum description) a chain of double
pendulums. The peculiar geometry of DNA produces rather involved
equations. The model is described by the Lagrangian density
\begin{eqnarray}\lb{lagraq}
&&\mathcal{L} = \frac12 \left\{ I \vth_t^2 - \om_t \vth_x^2 + r^2 [ m
\vth_t^2+\right. \nonumber\\
&&- \om_s \vth_x^2 + 2 (m \vth_t (\vphi_t + \vth_t) - \om_s
(\vphi_x + \vth_x) ) \cos \vphi \nonumber\\ && \left. + m (\vphi_t +
\vth_t)^2 - \om_s (\vphi_x + \vth_x)^2 ] \right\} +  \nonumber\\
&&+ 4 r^2 K_p [ \cos
\vth + \cos (\vphi + \vth ) - (1/2) \cos \vphi - (3/2) ] \ ,
\end{eqnarray} and we refrain from writing the Euler-Lagrange
equations for this Lagrangian; the reader is referred to \cite{CDGspeed}
for the physical meaning of the various parameters appearing in
$\mathcal{L}$.

Under the TWA, the Euler-Lagrange equations reduce to
\begin{eqnarray}
    \lb{jjj}
\mu  \, \vphi_{zz} &+& \mu  (1 +  \cos \vphi ) \, \theta_{zz}
 =\nonumber\\
 & &=- 4  K_p  \, \sin (\vphi + \theta )  -\mu \, \sin (
\vphi ) \, (\theta' )^2 \ + \ 2 K_p  \, \sin (\vphi)
 \ ; \nonumber\\
\mu \ (1 &+&  \cos \vphi ) \, \vphi_{zz} \ + \ [ (J /r^2) + 2\mu (1 +
 \cos \vphi )] \, \theta_{zz}\ =\nonumber\\
 & & =- 4  K_p (  \sin \theta + \sin (\vphi + \theta) ) +
\mu  \, \sin (\vphi) [ (\vphi_{z})^2 + 2 \vphi_{z} \theta_{z} ] \ ,
\end{eqnarray}
where $\mu := (m c^2 - \omega_s )$, $J
:= (I c^2 - \omega_t  )$, $\om_{t}=K_{t}\delta^{2}$,
$\om_{s}=K_{s}\delta^{2}$.

The system in invariant under the reflection symmetry $\theta\to
-\theta,\,\vphi\to -\vphi$.
In the Lagrangian (\ref{lagraq}) appear two different limit speeds, thus
space-time Lorentz symmetry is broken and we expect case $3)$ of
Sect. \ref{sect:tdc} to apply, i.e that the invariant submanifold
associated with the reflection symmetry  exists
only for fixed value of $c$.
The natural candidates as invariant
submanifolds of the dynamics are therefore $\theta=0$ and $\vphi=0$.
One can easily check that this is not the case for $\theta=0$,
whereas  setting $\vphi=0$ in Eq. (\ref{jjj}),  and dividing the second by a factor 4, we get

\beq \mu \, \theta_{zz} \ = \ - \, 2 \, K_{p} \, \sin\theta \ , \ \
\left(\frac{J}{4 r^{2}} + \mu\right) \, \theta_{zz} \ = \ - \, 2 \,  K_{p} \, \sin\theta \ .
\eeq
Thus $\vphi = 0$ is  an invariant manifold if  and only if
$J= 0$;  this fixes the speed of the TW to $c =\om_{t}/I$.

\subsection{Kinks in coupled chains of adsorbed atoms}
\label{sec:adatoms}

Let us now consider chains of adsorbed atoms (also called
``adatoms''), and in particular the interaction between kinks in
such chains \cite{KMad,BK}. In our notation, these are described
by the equations \cite{KMad}
\begin{eqnarray}
\vphi_{tt} \ - \ \vphi_{xx} &=& - \sin \vphi \ - \ \a \, \sin
(\vphi - \vpsi) \ + \ \b \, \vpsi_{xx} \ , \nonumber \\
\vpsi_{tt} \ - \ \vpsi_{xx} &=& - \sin \vpsi \ - \ \a \, \sin
(\vpsi - \vphi) \ + \ \b \, \vphi_{xx} \ . \label{eq:adatoms}
\end{eqnarray}
(These are also studied in \cite{KivCC} in the
case $\b = 0$.)
These correspond to the Lagrangian \beq\label{eq:lagadatoms} L \ =
\ \frac12 \left[ (\vphi_t^2 - \vphi_x^2) \ + \ (\vpsi_t^2 -
\vpsi_x^2) \right] \ - \ \b \, \vphi_x \, \vpsi_x \ + \ \left[
\cos (\vphi) + \cos (\vpsi) + \cos(\vphi - \vpsi) \right] \ . \eeq

The TWA produces the equations \begin{eqnarray*} (c^2 - 1)
\vphi_{zz} &=&
 \b \vpsi_{zz} - \sin \vphi - \a \sin (\vphi - \vpsi) \\
 (c^2 - 1) \vpsi_{zz} &=&  \b \vphi_{zz} \ - \ \sin \vpsi \ - \ \a
\sin (\vpsi -
 \vphi) \ , \end{eqnarray*}
which are written in the form
$$ M \ \Phi'' \ = \ F $$ by setting
$$ M \ = \ \pmatrix{c^2 - 1 & - \b \cr - \b & c^2 - 1 \cr} \ , \ \
\ F \ = \ \pmatrix{- \sin \vphi \ - \ \a \sin (\vphi - \vpsi ) \cr
- \sin \vpsi \ - \ \a \sin (\vpsi - \vphi ) \cr} \ ; $$ note that
$\det (M) = (c^2 - 1)^2 - \b^2$. Under the condition $c^2 \not= 1
\pm \b$, the equation is rewritten as
\begin{eqnarray*}
\vphi_{zz} &=& - \, \frac{ (1 - c^2) \, \sin \vphi \ + \ \a (1 +
\b - c^2 ) \sin (\vphi - \vpsi ) \ - \ \b \sin \vpsi }{\b^2 \ - \
(1-c^2)^2} \ , \\
\vpsi_{zz} &=& - \, \frac{ (1 - c^2) \, \sin \vpsi \ + \ \a (1 +
\b - c^2 ) \sin (\vpsi - \vphi ) \ - \ \b \sin \vphi }{\b^2 \ - \
(1-c^2)^2} \ . \end{eqnarray*}

This obviously admits two discrete symmetries:
$$ (\vphi , \vpsi) \to (\vpsi , \vphi ) \ \ \mathrm{and} \ \ (\vphi ,
\vpsi) \to (-
\vphi ,  - \vpsi ) \ . $$ The latter is of no use (the invariant
set it identifies is just $\vphi = \vpsi = 0$, which gives the
trivial solution), while the former suggest to pass to field
coordinates
$$ \eta \ := \ \frac{\vphi - \vpsi}{2} \ , \ \ \xi \ := \ \frac{\vphi
+ \vpsi}{2} \
. $$ In terms of these, the equations are rewritten as
\begin{eqnarray*}
\eta_{zz} &=& - \, \frac{\a \sin (2 \eta) \, + \, \sin (\eta) \,
\cos (\xi)}{(c^2 - 1) + \b} \ = \ - \, \sin (\eta) \ \frac{2 \a
\cos (\eta) \, + \, \cos (\xi)}{(c^2 - 1) + \b}\\
\xi_{zz} &=& - \, \sin (\xi) \ \frac{\cos (\eta)}{(c^2 - 1) - \b}
\ . \end{eqnarray*} We see immediately that both the submanifolds
identified by $\xi = 0$ and by $\eta = 0$ are invariant.

In the equation of motion  (\ref{eq:adatoms}) appears one single
limit speeds, which with the units used in
\eqref{eq:adatoms} is
$c_{1}=c_{2}=\bar c=1$, and a non vanishing kinetic coupling term
$\beta$, with $|\beta|\le1$. Thus space-time Lorentz symmetry is
broken only by  this coupling term, and case
$2)$ of Sect. \ref{sect:tdc}  applies. We
therefore expect  that the invariant submanifold associated with
the discrete symmetry  exists for every value of $c$ below a limit
value depending on $\beta$.  We will now show that this is
indeed the case.

For $\b = 0$ we would have a motion in an effective
potential
$$ W \ = \ - \frac{1}{c^2 - 1} \ \left[ \cos (\xi + \eta) \ + \
\cos (\xi - \eta) \ + \ \a \, \cos (2 \eta) \right] \ ; $$ the
analysis would be rather simple. The case $\b = 0$ corresponds to
not having the kinetic interaction term, see \eqref{eq:adatoms},
\eqref{eq:lagadatoms}.

By restricting to $\eta = 0$ and writing $h_+ = [c^2 -1 - \b
]^{-1}$ we obtain \beq \xi_{zz} \ = \ - \, h_+ \ \sin (\xi) \ ,
\eeq i.e. a standard sine-Gordon equation, which will support the
standard sine-Gordon solitons \cite{PD,Dun}.

Note however that we have $h_+ < 0$ if and only if $c^2 < 1 + \b$;
thus $\xi$-waves are possible if and only if the speed $|c|$ is
 \emph{smaller} than a critical value, $|c| < c_\xi = \sqrt{1 + \b}$.

By restricting instead to $\xi = 0$ and writing now $h_- = [(c^2
-1) + \b]^{-1} $ we get \beq \eta_{zz} \ = \  - \, h_- \ \sin
(\eta) \ [ 1 + 2 \a \cos (\eta) ] \ , \eeq i.e. a double
sine-Gordon equation, which will also support solitons
\cite{CaD}.

Note that here too $h_- < 0$ is not automatic: this is the case if
and only if $c^2 < 1 - \b$, i.e. again $\eta$-waves are possible
if and only if the speed is smaller than a limit  value, which is
now  $|c| < c_\eta = \sqrt{1 - \b}$.

Consistently with our discussion of the linear case, in order to
have  a real speed $c$ we must require $|\b|\le 1$ (corresponding
to the condition $|\ga|\le 1$ used in Sect. \ref{sec:exalin}).

It should also be noted that for $\b=0$ space-time Lorentz
invariance if fully preserved, i.e it is realized the case $1)$
described in Sect. \ref{sect:tdc}. The invariant submanifold
exists for every value of the soliton speed smaller than $\bar c =
1$.

\subsection{Coupled Josephson junctions}
\label{sec:CJJ}

A weakly coupled system of two long Josephson junctions is
described, in suitable units, by the equations \cite{KM}
\begin{eqnarray*}
\vphi_{tt} \ - \ \vphi_{xx} &=& - \sin (\vphi) \ - \ \a \,
\psi_{xx} \ - \ \s_1 \, \vphi_t \ + \ f_1 \ , \\
\mu^2 \, \vpsi_{tt} \ - \ \vpsi_{xx} &=& - \nu^2 \, \sin (\vpsi) \
- \ \a \, \vphi_{xx} \ - \ \s_2 \, \mu \nu \psi_t \ + \ \nu^2 f_2
\ ; \end{eqnarray*} we refer to \cite{KM} for the physical meaning
of the different constants. Note that here $\s_i$ describe
dissipation effects, and $f_i$ are external forcing needed to
counterbalance dissipation. Considering the idealized case where
dissipation is zero (and hence setting to zero also the external
forcing terms $f_i$), we are reduced to
\begin{eqnarray}
\vphi_{tt} \ - \ \vphi_{xx} &=& - \sin (\vphi) \ - \ \a \,
\psi_{xx} \ , \nonumber \\
\mu^2 \, \vpsi_{tt} \ - \ \vpsi_{xx} &=& - \nu^2 \, \sin (\vpsi) \
- \ \a \, \vphi_{xx} \ . \label{eq:CJJ} \end{eqnarray} This
corresponds to a potential $ V = - \, [\cos (\vphi ) + \nu^2 \cos
(\psi )]$.

The TWA reduces \eqref{eq:CJJ} to
\begin{eqnarray}
(c^2 -1) \, \vphi_{zz} \ + \ \a \, \vpsi_{zz} &=& - \sin (\vphi) \ ,
\nonumber \\
(\mu^2 c^2 - 1) \, \vpsi_{zz} \ + \ \a \, \vphi_{zz} &=& - \nu^2
\, \sin ( \vpsi) \ ; \label{eq:CJJ2}
\end{eqnarray}
This is written as $M \Phi_{zz} = F$ if we define
$$ M \ = \ \pmatrix{c^2 - 1 & \a \cr \a & \mu^2 c^2 - 1 \cr} \ , \
\ \Phi = \pmatrix{\vphi \cr \vpsi \cr} \ , \ \ F = \pmatrix{- \sin
\vphi \cr - \nu^2 \sin ( \vpsi ) \cr} \ .
$$ Provided $\det (M) := [(c^2 -1 ) (\mu^2 c^2 - 1) - \a^2] \not= 0$,
and writing then $h = [\det (M)]^{-1}$, the equations
\eqref{eq:CJJ2} are rewritten as $\Phi_{zz} = M^{-1} F$, i.e. as
\begin{eqnarray}
\vphi_{zz} &=& - \, h \ \left[ (1 - \mu^2 c^2) \sin (\vphi) \ + \
\a \, \nu^2 \sin (\vpsi ) \right]  \nonumber \\
\vpsi_{zz} &=& - \, h \ \left[ \nu^2 (1 - c^2) \sin (\vpsi) \ + \
\a \, \sin (\vphi) \right]. \label{eq:CJJ3}
\end{eqnarray}

We would like to determine one-dimensional invariant submanifolds
for the dynamics \eqref{eq:CJJ3}.  We look for discrete
symmetries of the model.
The system (\ref{eq:CJJ}) is invariant under the reflection
symmetry $\vphi \to -\vphi,\, \psi\to -\psi$. The most natural
candidates for invariant submanifolds are therefore $\vphi=0$ and
$\psi=0$. These are, however, of no use, because correspond to
trivial solutions of the field equations.
Thus  in general the determination  of the invariant submanifold is a  hard
problem, but we can look for \emph{linear} invariant submanifolds.

That is we will look for linear combinations of the fields, $\xi =
d_1 \vphi + d_2 \vpsi$ and $\eta = d_3 \vphi + d_4 \vpsi$, such
that the submanifolds $\xi=0$ and $\eta =0$ are invariant under
\eqref{eq:CJJ3}. The inverse transformation can be written
as\footnote{Writing $\rho = (k_1 k_3 - k_2 k_4)^{-1}$, the
relation between the $d_i$ and the $k_i$ coefficients is given by
$d_1 =- k_4 \rho$, $d_2 = k_1 \rho$, $d_3 = k_3 \rho$, $d_4 = -
k_2 \rho$.}
$$ \vphi = k_1 \, \eta \ + \ k_2 \, \xi \ , \ \ \vpsi \ = \ k_3 \,
\eta \ + \ k_4 \, \xi \ ; $$ in this way, and setting $$ H \ = \
\frac{h}{k_1 k_3 - k_2 k_4}  \ = \ h \ \rho \ , $$  the equations \eqref{eq:CJJ3}
are written as
\begin{eqnarray*} \xi_{zz} &=& - \, H \ \left[ [(\a k_1 - k_4 (1 -
c^2 \mu^2)] \sin (k_1 \eta + k_2 \xi)
\right.  \\ & & \ \ \ \left. \ + \ [(1 - c^2) k_1 - \a k_4] \nu^2
\sin (k_4 \eta + k_3 \xi)
\right] \\
\eta_{zz} &=& - \, H \ \left[ [(1 - c^2 \mu^2) k_3 - \a k_2 ] \sin
(k_1 \eta + k_2 \xi ) \right. \\ & & \ \ \ \left. \ - \ [(1 - c^2)
k_2 + \a k_3] \nu^2 \sin (k_4 \eta + k_3 \xi ) \right] \ .
\end{eqnarray*}

On the line $\eta = 0$ these reduce to
\begin{eqnarray*} \xi_{zz} &=& - \, H \ \left[ [(\a k_1 - k_4 (1 -
c^2 \mu^2)] \sin (k_2 \xi)
\ + \ [(1 - c^2) k_1 - \a k_4] \nu^2 \sin (k_3 \xi)
\right] \\
\eta_{zz} &=& - \, H \ \left[ [(1 - c^2 \mu^2) k_3 - \a k_2 ] \sin
(k_2 \xi ) \ - \ [(1 - c^2) k_2 + \a k_3] \nu^2 \sin (k_3 \xi )
\right] \ ; \end{eqnarray*} thus the line $\eta = 0$ is invariant
if and only if \beq\label{eq:CJJxi} (1 - c^2 \mu^2) k_3 - \a k_2 \
= \ 0 \ \ \mathrm{and} \ \ \ (1 - c^2) k_2 + \a k_3 \ = \ 0 \ .
\eeq

This can be cast in the form of a matrix equation $Q_\xi K_\xi =
0$, where we have defined
$$ Q_\xi \ = \ \pmatrix{- \a & (1 - c^2 \mu^2) \cr (1 - c^2) & \a
\cr} \ , \ \ K_\xi \ = \ \pmatrix{k_2 \cr k_3 \cr} \ . $$ The
solution exists provided $\det (Q_\xi ) = 0$, and this condition
considered as an equation for $c$ provides
$$ c^2 \ = \ c^2_{(\xi,\pm)} \ = \ \frac{(1 + \mu^2 ) \ \pm \
\sqrt{(1 - \mu^2)^2 + 4 \a^2 \mu^2}}{2 \ \mu^2} \ . $$

Notice that, in accordance with our results in the linear case, in
order to have  a real speed $c$ we must require $|\a|\le 1$
(corresponding to the condition $|\ga|\le 1$ used in Sect.
\ref{sec:exalin}).

Denoting by $Q_{(\xi,\pm)}$ the matrix $Q_\xi$ with $c^2 =
c^2_{(\xi,\pm)}$, the kernel of $Q_{(\xi,\pm)}$ is spanned
respectively by $$  v_{(\xi,\pm)} \ = \ \pmatrix{(1 - \mu^2) \,
\pm \, \sqrt{(1-\mu^2)^2 + 4 \a^2 \mu^2} \cr 2 \a \cr} \ .
$$

One can check that with these choices for $c$ and for $k_1,k_4$
the line $\eta=0$ is invariant, and the evolution of $\xi$ is
governed by an equation of the form
$$ \xi_{zz} \ = \ p_1 \, \sin ( 2 \a \xi) \ + \
p_2 \, \sin (\om_{\xi,\pm} \xi ) \ , $$ where $p_i$ are
coefficients depending on $(k_1,k_4)$, whose explicit expression
can be easily computed but is not interesting here, and
$$ \om_{(\xi,\pm)} = - (1 + \mu^2) \ \pm \ \sqrt{(1- \mu^2)^2 + 4 \a^2
\mu^2} \ . $$ One could analyze in the same way the invariance of
the line $\xi=0$. Note however that there is no physical
difference between the two fields, as they are just generic linear
combinations of $\vphi$ and $\vpsi$; thus the $\xi=0$ case will
reproduce -- with an exchange of roles between $\xi$ and $\eta$ --
the same results obtained for $\eta=0$.

Note that in the case analyzed here the Lorentz symmetry is broken
by the presence of two  different limit speeds in  the field
equations (\ref{eq:CJJ}). Thus, case $3)$ of Sect. \ref{sect:tdc}
applies. The invariant submanifold associated with the discrete
symmetry  exists  only  for selected values of the speed $c$.

\subsection{Modified Katsura model}
\label{sec:Katsura}

Our last example is a ``two speeds of light'' generalization of
the model of  Katsura \cite{Kat}; this  has been proposed to
describe the coupling of  magnetic and ferroelectric domain walls
\cite{Kat,FK}. The model we consider is described by the
Lagrangian (for simplicity we set in the model of Ref. \cite{Kat}
the coupling constant  $\la=\mu=1$) \beq\lb{kat} {\cal{L}}=
\frac{1}{2}\left(\frac{\vphi_{t}^{2}}{c_{1}^{2}}- \vphi_{x}^{2}+
\frac{\psi_{t}^{2}}{c_{2}^{2}}- \psi_{x}^{2}\right)-V(\vphi,\psi)
\ , \ \ \ V(\vphi,\psi)= \frac{\vphi^{4}}{4}- \vphi \cos\beta\psi.
\eeq

The Lagrangian has the discrete symmetry $\psi\to -\psi$, so that
the natural candidate for the invariant submanifold for the
dynamical system one obtains after the TWA  is $\psi=0$. Indeed
such invariant manifold exists for every value of the parameters
and the reduced dynamics is described by $ \vphi_{zz} =
\vphi^{3}-1$.
However, this equation does not support solitonic
solutions because the potential  does not have the required minima
structure.

The model also has another reflection symmetry, described by
$$ \vphi \to - \vphi \ , \ \ \vpsi \to \frac{\pi}{\b} - \vpsi \ . $$
If we look for analytic functions invariant under such a reflection,
they are necessarily built as (algebraic) functions of the basic
invariants
$$ \vphi^2 \ , \ \ \cos^2 (\b \psi) \ , \ \ \vphi \, \cos \b \psi \ .
$$

In fact, another invariant manifold for the dynamics  can be found
using the results of Ref. \cite{Kat}, where exact kink solutions
have been found for $\beta^{2} = 1/2$ using the ansatz
\beq\lb{ans} \vphi= \cos\beta \psi. \eeq One can easily show that
Eq. (\ref{ans}) with $ \beta^{2} \neq 1/2$ determines an invariant
manifold for the equivalent mechanical system describing the TW
dynamics stemming from \eqref{kat}, if the TW speed is fixed by
\beq\lb{kk1} c \ = \ c_{2} \, c_{1} \
\sqrt{\frac{1-2\beta^{2}}{c_{1}^{2}-2\beta^{2}c_{2}^{2}}}. \eeq

Note that again we have two limiting speeds in the Lagrangian
(\ref{kat}); thus the model falls in case 3) of section
\ref{sect:tdc}, and we do indeed have the behavior predicted
there.  On the other hand,  in the case $c_{1}=c_{2}$ and
$\beta^{2} = 1/2$,  consistently with the results of Katsura
\cite{Kat},   equation (\ref{ans}) describe an invariant manifold
of the dynamics \emph{for every value of $c$}; thus in this special
case the system falls in case 1) of the classification given in
Sect. \ref{sect:tdc}.

\section{Discussion. Field theoretical considerations}
\label{sec:discFT}

This paper  focused on the description of TW dynamics in
macroscopic and mesoscopic systems. This is particular evident in
the examples we have chosen in Sects. \ref{sec:exalin} and
\ref{sec:exanonlin} to elucidate our mechanism. They represent
macroscopic or mesoscopic models for molecular, biological or
condensed matter systems.

Nevertheless, at least some of the results of these paper have a much
broader relevance and can be applied, generically, to any
two-dimensional (2D) field  theory. This is particularly true
for  the first part of  paper, which  concerned linear
dynamics.

An important point in this context is the presence in the action
for the 2D  field theory (\ref{eq:lag}) of $N$ different ``speeds
of light'' $c_{i}$,  see Eq. \eqref{eq:speed}. This is a natural condition
for macroscopic and mesoscopic  systems, where different modes
propagating in a medium experience different effective physical
parameters $\rho_{i},\k_{i}$ (elastic, optical, magnetic and so
on) of the media so that at the linear level their perturbations
propagate at different speeds. On the other hand a fundamental,
microscopic, 2D field theory has to be Lorentz invariant, thus
characterized by a single speed of light.

The case of a single  limit speed is a particular case of the
description of linear waves given  in the first part of this paper
(see e.g Eq. (\ref{top})). Thus, our results including those
related to synchronization and tuning of the wave speed hold true
for $c_{1}=c_{2}$.

From the field theoretical point of view a particularly important
case  is that  of a non interacting  theory. When $V =\ga=0$ and
$c_{i}=\bar c$ the theory (\ref{eq:lag}) is a 2D conformal field
theory (CFT) . It describes CFTs with $N$ scalar fields (i.e. CFTs
with central charge $N$), which play a fundamental role in several
contexts such as string theory \cite{Pol}, critical points of
phase transitions \cite{Cardy}, microscopic explanation of black
hole entropy \cite{Carlip}, just to mention only few of them. The
conformal invariance of  the Lagrangian (\ref{eq:lag}) is fully
evident by passing to light cone coordinates $x_{\pm}= x\pm c t$.
In this coordinates the theory is explicitly invariant under the
action of 2D-diffeomorphisms (the conformal group in 2D) and  the
field equations read  $\partial_{x_+}
\partial_{x_-} \vphi = 0$, whose general solution is a generic
combination of right and left moving TW $\vphi  = f(x_{+}) +
f(x_{-})$. The conformal invariance is a general feature of the
massless case. In fact it is  also evident from Eq.
(\ref{eq:linTWV0}), which does not fix the dependence on $z$ of
the fields $\vphi(z),\,\psi(z)$ but just gives a linear relation
between the two fields.

As a final remark  we note an analogy between the
synchronization mechanism for TW  in the linear regime  we have
found in the first part of this paper, and Quantum Mechanics; this goes as follows.

Let us consider for simplicity the field equations (\ref{tweq})
with vanishing potential (our considerations can be easily
extended to the $V\neq 0$ case). The system can be described in
two  different frames of the field space $\vphi,\psi$: a
frame in which the two fields decouple completely
i.e. in which the kinetic matrices are diagonal; and a frame in which the
two fields interact with  the interaction term $\ga$.  In the
first frame TW for the two fields  can propagate independently,
hence with different speeds $c_{+}$ and $c_{-}$ given by Eq.
(\ref{kgh}). Owing to the interaction, in the second frame
the $\vphi-$ and $\vpsi$-waves are forced to travel with the same
speed (synchronization).

This relationship between TW waves in the two  frames bears  a
strong analogy with the eigenstates of two non-commuting quantum
mechanical operators, acting on a 2D Hilbert spaces and associated
to two non compatible physical observables.

\section{Conclusions}

In previous work, dealing with a concrete model for the nonlinear
dynamics of DNA \cite{CDG1}, we  observed a peculiar mechanism
which fixed the speed of solitons \cite{CDGspeed}. In this paper
we investigated if this mechanism could extend to a more general
class of equations, answering this question in the positive. More
generally, we have considered systems of coupled wave equations;
when decoupled each of them is Lorentz invariant with a limit
speed $c_i$, with possibly different limit speeds and we have
studied how the speed of travelling waves is affected by the
coupling.

In particular we have considered systems admitting a variational
(Lagrangian) description, and in which the coupling between the
different equations could involve a kinetic term, see Eq.
\eqref{eq:lag}.

We observed that the general Lagrangian under consideration here,
(see Eq. \eqref{eq:lag}) could fall in different classes according to
its properties under Lorentz transformations. That is, Lorentz
invariance could be unbroken (case 1), in which case there is of
course no speed selection mechanism; it could be broken, albeit
the coupled equations admit the same limit speed, by coupling
terms (case 2), in which case there is still no speed selection
but the limit speed can be changed due to the coupling; or finally
the Lorentz invariance can be broken by the presence of different
limit speeds \emph{and} kinetic coupling terms (case 3), in which
case we have a full selection of the speed of travelling wave
solutions, which can only take a finite -- and rather small, being
limited by the number of coupled equations -- set of values.

In the latter -- and more interesting -- case, simple travelling
wave solutions can be described as dynamically invariant
one-dimensional manifolds for a mechanical system (parametrically
dependent on the speed $c$ of travelling waves) associated to the
system of PDEs under study, joining two extremal points for an
effective potential which comply with limit conditions dictated by
the finite energy condition for the PDEs system. Such manifolds
only exist for special values of $c$, and this ignites the speed
selection mechanism.

We have then validated our general discussion by a number of
physically significant examples; in particular, in Sects.
\ref{sec:DPC}, \ref{sec:CJJ} and \ref{sec:Katsura} we have shown
that the speed selection mechanism here considered is present in
double pendulum chains, coupled Josephson junctions, and in a
modified Katsura model for the coupling of magnetic and
ferroelectric domain walls.

It is interesting that the mechanism described here -- and which
has a partial counterpart for linear equations as well --  has
some points of contact with general field-theoretic questions, as
discussed in Sect. \ref{sec:discFT}.

In the Appendices, we give a closer look at some group-theoretical
questions (Appendix \ref{sec:group}); and argue that, albeit here
we worked specifically with hyperbolic PDEs, the same mechanism
can be at work in some type of parabolic equations (Appendix
\ref{sec:parabolic}).

\section*{Acknowledgements}

We thank  M. Tarallo and S. Walcher for useful comments. The work
of GG is partially supported by MIUR-PRIN program under project
2010-JJ4KPA.

\vfill\eject

\appendix
\section{Lorentz symmetry and group theoretical considerations}
\label{sec:group}

In Sect. \ref{sect:nfg} we have  noted that in the Lagrangian
(\ref{eq:lag}) the usual space-time Lorentz symmetry is
explicitly broken by  the presence  of several limiting speeds
$c_{i}$. However,
 there is some remnant of the Lorentz invariance; this can be
described as follows.

The Lorentz transformation \beq\label{eq:Lorentz} x \to \wt{x} \ =
\ \frac{(x - v t)}{\sqrt{1 - v^2/c^2}} \ , \ \ t \to \wt{t} \ = \
\frac{( t - (v/c^2) x)}{\sqrt{1 - v^2/c^2}}  \eeq is generated by
the vector field
$$ X \ = \ - \frac{x}{c^2} \ \frac{\pa}{\pa t} \ - \ t \
\frac{\pa}{\pa x} \ . $$ The \emph{evolutionary representative}
\cite{Olv,Ste,KrV,CGbook} for this is the (generalized) vector
field
$$ X_{ev} \ = \ \left( \frac{x}{c^2} \, u_t \ + \ t \, u_x \right) \
\frac{\pa}{\pa u} \ , $$
where $u$ is the dependent variable (the physical field).

Thus if we consider the Lorentz symmetry with several limit speeds $c_i$
acting on the field $\phi^i$, this is generated in the
evolutionary representation by
$$ X_{ev}^i \ = \ \left( \frac{x}{c_i^2} \, \phi^a_t \ + \ t \,
\phi^a_x \right) \
\frac{\pa}{\pa \phi^a} \ . $$ If now we consider a generalized
vector field
$$ X_L \ = \ X_{ev}^1 + ... + X_{ev}^N \ , $$
which (unless $c^2_1 = ... = c^2_N$) will \emph{not} be the
evolutionary representative of any Lie-point vector field acting
on the space of the $(x,t;\phi^1 , ... , \phi^N )$ variables, this
will be a symmetry for the Lagrangian $\L_1 + ... + \L_N$, where
we define the partial Lagrangians as
$$ \L_i \ = \ \frac12 \ \left[ \rho_i^2 \, (\phi^i_t)^2 \ - \
\k_i^2 \, (\phi^i_x)^2  \right] \ , $$ and for the corresponding
field equations $ \phi^i_{tt} - c^2_i \phi^i_{xx} = 0$ as
well.\footnote{Actually, here only derivatives of first (for the
Lagrangian) and second (for the field equations) order matter; so
we could as well consider only transformations of the field
derivatives, e.g. prescribing these are undergoing a hyperbolic
rotation irrespective of any transformation on the field
themselves. In this way one would be led to consider
``hidden symmetries'' \cite{Cariglia}.}

Note however that this will \emph{not} leave the interaction
Lagrangian, nor the corresponding interaction terms in the field
equations, invariant.

The symmetry properties of the Lagrangian and field equations
described above can be also understood in terms of the
commutation relations (\ref{eq:Qcomm}) of the matrices
$Q^{(t)},Q^{(x)}$ defined  in (\ref{eq:Qmat}).  We have a symmetry
of the Lagrangian when the matrices can be diagonalized
simultaneously, i.e  when $\ga=0$ (corresponding to absence of
coupling in the kinetic sector) and  $c_{i}=\bar c$, i.e  the same limit
 for the TW of  different fields.

The $N$-field components  TW will  belong to the representation of
the Lorentz symmetry generated by the generalized vector field
$X_{L}$. More specifically they will transform via reducible
representations
$$ T \ = \ T_1 \oplus T_2 \oplus ... \oplus T_N \ , $$ where each
$T_a$ is a two-dimensional representation made of the Lorentz
group. The TW  solutions of the model with vanishing, respectively
non-vanishing,  potential belong to massless, respectively
massive, representations of the Lorenz group.

\section{Extension to  parabolic  equations}
\label{sec:parabolic}

We have so far considered coupled \emph{hyperbolic} equations; on the other
hand the treatment of the linear case is based on basic algebraic
facts and does not involve hyperbolicity; it thus appears that our
discussion can be extended to more general (time-evolution)
equations, and in particular to \emph{parabolic} ones.

Here, we will not discuss the general case but only consider, for
the sake of concreteness, TW dynamics described by coupled linear
Schr\"odinger equations
(LSE); this can be thought as emerging from the linearization of
the nonlinear Schr\"odinger equations (NLS) \beq \label{eq:NLS} i
\, \psi_t \ = \ - \,\frac{1}{2}\psi_{xx} \ + \ \k \, |\psi |^2 \,
\psi  \ . \eeq This is known to describe,  among others, optical
solitons and, upon introducing a potential term $V(x)\psi$,
Bose-Einstein condensates. As usual we will consider the
$N=2$ case, the generalization to the $N$-fields case
being performed along the lines described in Sect. \ref{sect:nfg}.

Let us consider the system of two coupled LSE,
\bea\lb{lse}
\frac{i}{c_{1}} \, \vphi_{t} \ +  \ \frac{1}{2} \, \vphi_{xx} \ - \ \frac{\ga}{2} \,
\psi_{xx} &=&
- \, \frac{\pa V}{\pa \vphi} \ ,\nonumber \\
\frac{i}{c_{2}} \, \psi_{t} \ + \ \frac{1}{2} \,  \psi_{xx} \ - \ \frac{\ga}{2} \,
\vphi_{xx} &=& - \, \frac{\pa V}{\pa \psi} \ , \label{eq:EL1}
\eea
where $c_{i},\ga$ are some parameters and now the fields $\vphi,\psi$
are
generically complex. In analogy  with the hyperbolic
case we have introduced  a
quadratic potential of the form  given by Eq. \eqref{pot1}.
As in the hyperbolic case,  to solve this equations we pass
to  the Fourier transforms.

In the decoupled case $\ga=\la=0$ we get two  dispersion relations
 $\om_i (q),\, i=1,2$
and the related phase velocities $v^{(p)}_\pm = d\om_\pm /dq$,
\beq
\om_i = \frac{1}{2}c_{i}\left(q^{2}-\mu_{i}^{2}\right) \ , \ \
v^{(p)}_{i}= c_{i}q \ ; \eeq
thus $c_{i}$ characterizes the phase velocity of the wave
packet.

In the generic coupled case $\ga\neq0,V\neq0$, we get dispersion
relations similar to
(\ref{eq:DRgen}),  i.e.
\bea\label{eq:DRgenls}
&&\om_\pm =
\frac{1}{2}\left[\frac{1}{2}(c_{1}+c_{2})q^{2}-(c_{1}\mu_{1}^{2}+c_{2}\mu_{2}^{2})
\pm \sqrt{\mathcal{P} (q^2)} \right],\\
 &&\mathcal{P} (q^2)
 = \frac{1}{4}(c_{1}+c_{2})^{2}q^{4}
 -(c_{1}\mu_{1}^{2}+
 c_{2}\mu_{2}^{2})(c_{1}+c_{2})q^{2}+ \nonumber \\
  &&+(c_{1}\mu_{1}^{2}+c_{2}\mu_{2}^{2})^{2}
-c_{1}c_{2}(q^{2}-2\mu_{1}^{2})(q^{2}-2\mu_{2}^{2})+c_{1}c_{2}(q^{2}\ga+\la)^{2}.
\end{eqnarray}

 Notice also that in the case of vanishing potential we get
the dispersion relations
\beq\lb{drls}
\om_\pm \ = \ \frac{1}{2} \ c_{\pm}q^{2} \ ,
\eeq
where $c_{\pm}$ has exactly the same  form (\ref{kgh}) found
in the hyperbolic case with squared-velocities replaced by the
parameters $c_{i}$ :
\beq\label{eq:vzeropotls} c_{\pm} \ = \ \frac{1}{2}  \ \left[
c_{1}+c_{2}\ \pm \ (c_{1}-c_{2}) \sqrt{1+
\frac{4 \, \ga^{2} \, c_{1} \, c_{2}}{(c_{1}-c_{2})^{2}} }
\right] \ . \eeq

As in the hyperbolic case this equation tells us that we can tune
the phase velocity of the $\vphi$ and $\psi$  waves just by changing
the parameter $|\ga|$ in the range $[0,1]$, (see Eq. (\ref{ub})). In particular, the phase
velocity of a given phase in $\om_+$ ($\om_-$) changes from a
maximum (minimum) value when $\ga=0$, given by the smallest
(greatest) of $c_\pm$, to zero (a maximum value) when $|\ga|=1$.

\vfill\eject

\end{document}